\documentstyle[12pt,aasms4,psfig,rotate]{article}

\begin{document}

\title{ On the Stability of  Quasi-Equilibrium Self-Gravitating 
 Configurations  
in a Tidal Field}

\author{M.A. G\'omez-Flechoso\altaffilmark{1} \& R. Dom\'{\i}nguez-Tenreiro}
\affil{Dept. F\'{\i}sica Te\'orica, C-XI. Univ. Aut\'onoma de Madrid,
E-28049 Madrid}

\altaffiltext{1}{Present address: Observatoire de Gen\`eve, 
Ch. des Maillettes 51, Ch-1290 Sauverny (Switzerland)}

\begin{abstract}

The possibility that quasi-equilibrium self-gravitating
galaxy-like configurations exist in a tidal field is analyzed in this
paper.
More specifically, we address the question of how to predict initial
configurations modeling galaxies that are able to survive
environmental effects  in a dense environment for a Hubble time or so,
provided thay dynamical friction is neglected.
For simplicity, the configurations in the tidal 
field have been taken initially to be
spherically symmetric and to have an isotropic velocity dispersion
tensor (t-limited King spheres); they orbit inside
steady state, spherical halos, as those that presumably surround compact
galaxy groups and galaxy clusters. 
Both circular and eccentric orbits have been considered.
In both cases, the initial quasi-equilibrium configurations have been
built up taking into account the external tidal field produced 
by the halo. 
It modifies the escape velocity field of the configuration, compared with
isolated configurations.
The survival of the configurations as they orbit inside the halos
has been studied through N-body simulations.
As a general result, it has been found out that the bulk of the
models is conserved along 12.5 Gyears of evolution, 
and that the  low rates of mass
losses   they experience are consistent with those
expected  when the adiabatic protection hypothesis is at work.
So, solutions for  galaxy configurations in tidal quasi-equilibrium have
been found, showing that
tidal stripping 
in quiescent phases  does not seem to be very important, unless that  the 
density of the galaxy environment 
at its formation had been much lower than that  of the galaxy environment 
at the point of its orbit where the tidal perturbation is maximum.

\end{abstract}


\section{Introduction}

Dense halos of dark matter have first been detected in galaxy clusters
and, then, in many galaxy groups, 
through observations of X-ray diffuse emission of the gas component 
surrounding the member galaxies
(Boehringer 1997; Ponman et al. 1996).
Also, some small dwarf galaxies are known to be orbiting inside the 
dark halos of larger ones
(e.g. Mateo 1998).
Interactions with such dense environments can cause these galaxies 
several dynamical effects (i.e. tidal heating,  mass loss from tidal stripping,
energy loss from dynamical friction, among others) that could
result in severe modifications, relative to isolated
galaxies, of the evolutionary history of both, individual galaxies
and the systems they form.
In fact, some characteristic times and scales playing relevant roles
in astrophysical processes, such as the typical times for decay 
to the center of the halos, the rates of background enrichment by
processed gas and so on, could change appreciably, depending
on the characteristics of the environment
(e.g., Gunn \&
Gott 1972; Merritt 1985; Moore, Lake, \& Katz 1998).

The modelization of environmental interactions has been mainly carried out 
through N-body simulations of the evolution of a theoretical galaxy model
moving in an external field 
(Barnes 1985; Funato, Makino, \& Ebisuzaki 1993; Bode, Cohn,
\& Lugger 1993; Bode et al. 1994; Garc\'{\i}a-G\'omez, Athanassoula, \& Garijo
1996; Athanassoula, Makino, \& Bosma 1997 and references quoted therein). 
One of the main shortcomings of galaxy models appearing in the
literature is that, in most cases, galaxies are built up
as if they were {\it isolated}. However, to properly quantify 
the effects of environmental interactions, it would be more convenient
that the galaxy model, at the beginning of the simulation,
 takes into account the
external forces. Otherwise, it is difficult to disentangle which
effects are effectively due to interactions and which ones are
spurious, due to an incorrect choice of the initial galaxy model (see 
G\'omez-Flechoso \& Dom\'{\i}nguez-Tenreiro 2000, hereafter GD00,
for a discussion). In this paper, we will focus on the 
choice of the initial
galaxy model  and on how 
this choice  affects the later evolution of the galaxy
as it orbits inside a halo. 
The effects of the dynamical friction and
the interactions between galaxies in groups and clusters will be tackled
in forthcoming papers.
  
 King models (King 1965, 1966) provide a good framework to study 
equilibrium self-gravitating configurations.
They are based on distribution functions that depend on
the  potential, $\Phi$,  causing the forces felt by the constituent
particles. So,   if the body is isolated,
the potential that enters in King distribution functions is
$\Phi = \Phi_S $, where $\Phi_S$ is due to the
mass distribution of the 
self-gravitating configuration (hereafter, satellite); however, when
it moves through an external halo, then $\Phi = \Phi_S + \Phi_{ext}$
must be used instead, where
$\Phi_{ext}$ is the potential causing the external force seen 
by the constituent particles.
This force also determines  the limiting or tidal radius
of a spherical configuration, $r_{\rm t}$,
that can be defined as the asymptotic distance at which constituent
particles remain stably bound to the satellite.
In spite of this, most King models found in literature are constructed
on the assumption that $\Phi = \Phi_S $
and take $r_{\rm t}$ as a free parameter, even if the body is not
isolated (Meylan \& Heggie 1997; but see also Heggie \& Ramamani 1995).
The tidal radius is a fundamental parameter of King models describing
spherically symmetric satellites, when, as due, the effects 
of the external field are explicitly taken into account
(hereafter, t-limited King models).

The stability of self-gravitating configurations relative to tidal
perturbations is equivalent to the stability of the orbits of
its constituent particles. 
After the pioneering work by King (1962), the problem of tidal limitations
imposed on such configurations has been studied in models provided
by the circular and elliptical restricted three body problems
(Keenan \& Innanen 1975; Jefferys 1976; Keenan 1981a and 1981b).
When the configuration is on a circular orbit, the problem can be worked out 
in some detail, as the equations of motion of their constituent
particles have one integral of motion, the Jacobi integral,
$E_J$, that can be used to define zero velocity surfaces in the
configuration space.
Stability against escape for a given orbit is assumed when the corresponding
zero velocity surface is closed (Spitzer 1987;
note that the zero velocity surface can be open
and, nevertheless, the particle does not escape over a given time interval).
And so, the limiting radius, 
$r_{\rm t}$, has been defined in terms of the distance, $x_e$, from the
satellite center  to the inner Lagrange point of the potential
$\Phi = \Phi_S + \Phi_{ext}$. However, in a tidal field 
zero velocity surfaces are
not spherically symmetric and, then, an ambiguity arises when one intends to
determine the radius of a spherically symmetric body embedded in a tidal
field that has not this symmetry. So, different authors 
 make different choices. King (1962) takes $r_{\rm t} = x_e$, while
Keenan (1981b) suggests that $r_{\rm t} = 2x_e/3$ is preferable
(see also Innanen, Harris, \& Webbink 1983; Spitzer 1987; Lee 1990; Heggie \&
Ramamani 1995);
these choices correspond to  two semiaxes of the
Roche surface.

No integral of motion exists when the satellite moves along a non
circular orbit, as the intensity of the tidal field changes 
periodically with time, being maximum at pericentric passage.
The question then arises of whether or not 
the energies of most constituent particles
 are  conserved to a good
approximation so that  {\it local}
(i.e., depending on the satellite orbital phase)
tidal  radii could be defined,
as in the circular motion case, that would translate into
satellites in tidal {\it quasi}-equilibrium
(note that, in the strict sense, {\it equilibrium} configurations
do not exist in a tidal field, because some degree of mass losses 
can never be avoided).
We see that more complications are added 
on the theoretical side to the ambiguities appearing
in the circular case. Observational data on the limiting radii of 
globular clusters cannot clarify the situation either, as no clear 
conclusion
has still been  reached on their possible dependence  
on the cluster orbital phase (Oh \& Lin 1992; 
Oh, Lin, \& Aarseth 1992; Meziane \& Colin 1996;
Brosche, Odenkirchen, \& Geffert 1999). In any case,
results on tidal equilibrium for globular clusters, where two-body
heat conduction plays an important role, could be not valid for
galaxy-like configurations, that are essentially collisionless systems.

 The purpose of this paper is to deepen into the understanding
of tidal quasi-equilibrium for self-gravitating galaxy-like configurations,
in particular to predict initial configurations for galaxy 
models which will survive environmental effects along a Hubble time or so.
The possibility to predict self-gravitating spherical collisionless
configurations in tidal quasi-equilibrium has been tested  
through Montecarlo realizations of the t-limited
King models. They are left to evolve in the corresponding external
field, which has been described by an analytical expression. 
Various possibilities have been explored about the
matching of the internal and external field of forces at the limiting
radius of the galaxy model.
Most of the previous works on tidal quasi-equilibrium
referred to globular clusters moving in a galactic potential.
As here we are mainly concerned with the influence of dense
environments (i.e., halos) on galaxy evolution, the halo density profiles
used in our test are those of dark halos appearing in N-body
simulations of hierarchically clustering universes (Navarro, Frenk, \& 
White 1996), and the parameters of the galaxy models correspond
to those of typical ellipticals.

The paper is organized as follows: in $\S$\ref{21} we give the general
expression for the tidal field caused by a spherical static halo in the
harmonic approximation.
In $\S$\ref{proforb} we specify the models and parameters of halos, 
orbits and galaxies used to make our study.
The results of this study are presented in $\S$\ref{results}.
Finally, in $\S$\ref{sumcon}, the summary and conclusions of the work are
given.

\section{The Tidal Field Caused By A Spherical Static Halo: General 
Expression}
\label{21}

As a first step to build up quasi-equilibrium
{\it initial} configurations in a tidal field,
in this Section we derive the general expression for the tidal field
caused by a spherical halo in the harmonic approximation.
Let us consider a satellite of mass $M_S$ distributed according
with a density profile $\rho_S(\mbox{\boldmath $r$}, t)$ embedded in a static,
spherically symmetric extended halo of total mass $M_H$ and density profile
$\rho_H(R)$. The satellite is assumed to move on 
an orbit characterized by energy $E_H$ and orbital angular
momentum per unit mass $\mbox{\boldmath $L$}_H$, relative to an inertial system
of 
reference. Let $\mbox{\boldmath $R$}_S(t)$ and $\mbox{\boldmath $V$}_S(t)$ be
the 
instantaneous position and velocity vectors of the
center of mass of the satellite, relative to an inertial
system, $S_O$, with origin at the center of potential, $O$, of 
the halo. Relative to the center of mass of 
the system formed by both, the satellite and the halo, the 
center of potential, $O$, and the center of mass of the 
satellite have position vectors

\begin{eqnarray}
\mbox{\boldmath $R'$}_O &=& \frac{-M_S/M_H}{1+M_S/M_H} \mbox{\boldmath $R$}_S =
{\cal
O}(M_S/M_H)\nonumber \\
\mbox{\boldmath $R'$}_S &=& \frac{1}{1+M_S/M_H} \mbox{\boldmath $R$}_S =
\mbox{\boldmath $R$}_S + {\cal O}(M_S/M_H)
\end{eqnarray}

Neglecting terms in $M_S/M_H$, the combined potential of the halo and
the satellite has spherical symmetry, and so, $\mbox{\boldmath $L$}_H$ is
conserved
in $S_O$. The satellite moves around the point $O$ with an 
instantaneous angular frequency $\mbox{\boldmath $\Omega$}(t) = \mbox{\boldmath
$L$}_H/R^2_S(t)$
(which is constant for circular orbits). In a  coordinate
system, $S_{\Omega}$, that rotates at an angular speed $\mbox{\boldmath
$\Omega$}(t)$ 
with respect to $S_O$, the equation of motion for the mass center of the
satellite is

\begin{equation}
\frac{d^2 \mbox{\boldmath $R$}_S}{dt^2}= -\left[\mbox{\boldmath $\dot{\Omega}$}
\times \mbox{\boldmath $R$}_S +
2 \mbox{\boldmath $\Omega$} \times \mbox{\boldmath $\dot{R}$}_S +
\mbox{\boldmath $\Omega$} \times (\mbox{\boldmath $\Omega$}
\times \mbox{\boldmath $R$}_S) \right] - \nabla \Phi_H(\mbox{\boldmath $R$}_S)
\label{eqor}
\end{equation}

\noindent
where the first, second and third terms on the r.h.s. of Eq. (\ref{eqor})
are the inertial force of the rotation, the Coriolis force and the 
centrifugal force at $\mbox{\boldmath $R$}_S$ and the last term is the force
caused
by the mass distribution of the halo at point $\mbox{\boldmath $R$}_S$.

Let us now consider a bound particle P belonging to the
satellite, whose position relative to $S$ is $\mbox{\boldmath $r$}$ and
the relative to $O$ is $\mbox{\boldmath $R$}_P$ ($\mbox{\boldmath $R$}_P
=\mbox{\boldmath $R$}_S + \mbox{\boldmath $r$}$).
The equation of motion for P in $S_{\Omega}$ is

\begin{equation}
\frac{d^2 \mbox{\boldmath $R$}_P}{dt^2}= -\left[\mbox{\boldmath $\dot{\Omega}$}
\times \mbox{\boldmath $R$}_P +
2 \mbox{\boldmath $\Omega$} \times \mbox{\boldmath $\dot{R}$}_P +
\mbox{\boldmath $\Omega$} \times (\mbox{\boldmath $\Omega$}
\times \mbox{\boldmath $R$}_P) \right] - \nabla \Phi_H(\mbox{\boldmath $R$}_P)
- \nabla_r \Phi_S(\mbox{\boldmath $R$}_P)
\label{eqor1}
\end{equation}

\noindent
where the last term is the force on P caused by the mass distribution of the 
satellite.

 From Eqs. (\ref{eqor}) and (\ref{eqor1}) the equation of motion of P in 
the  coordinate system, $S_S$, centered at S and that
rotates with instantaneous angular speed $\mbox{\boldmath $\Omega$}(t)$ with 
respect to $S_O$ is

\begin{equation}
\frac{d^2 \mbox{\boldmath $r$}}{dt^2}= -\left[\mbox{\boldmath $\dot{\Omega}$}
\times \mbox{\boldmath $r$} +
2 \mbox{\boldmath $\Omega$} \times \mbox{\boldmath $\dot{r}$} + \mbox{\boldmath
$\Omega$} \times (\mbox{\boldmath $\Omega$}
\times \mbox{\boldmath $r$}) \right] - \nabla_r \Phi_S(\mbox{\boldmath $r$}) 
- \nabla \Phi_H(\mbox{\boldmath $R$}_P) + \nabla \Phi_H(\mbox{\boldmath $R$}_S)
\label{eqor2}
\end{equation}

and a series development around $\mbox{\boldmath $R$}_S$ gives, at first order
in 
$r/R$:

\begin{equation}
\frac{d^2 \mbox{\boldmath $r$}}{dt^2} = - [\mbox{\boldmath $\dot{\Omega}$}
\times \mbox{\boldmath $r$} + 2 \mbox{\boldmath $\Omega$}
\times \mbox{\boldmath $\dot{r}$}] - \nabla_r \Phi_S(\mbox{\boldmath $r$}) -
\nabla_r \Phi^{\rm tidal}(\mbox{\boldmath $r$})
\label{eqorfin}
\end{equation}

where

\begin{equation}
\Phi^{\rm tidal}(\mbox{\boldmath $r$};\mbox{\boldmath $R$}_S,\mbox{\boldmath
$\Omega$}) = \beta r^2 +
(\alpha-\beta)\left(\frac{\mbox{\boldmath $r$}\cdot \mbox{\boldmath
$R$}_S}{R_S}\right)^2 +
(\gamma-\beta)\left(\frac{\mbox{\boldmath $r$}\cdot \mbox{\boldmath
$\Omega$}}{\Omega}\right)^2
\label{phiterm}
\end{equation}

and

\begin{eqnarray}
\label{coef}
\alpha & = & \frac{1}{2} (\Phi''_H(R_S) -\Omega^2) = 2 \pi G ( \rho_H(R_S) -
\frac{2}{3} \overline{\rho_H}(R_S))-\Omega^2/2\nonumber\\
\beta & = & \frac{1}{2}(\Phi'_H(R_S)/R_S - \Omega^2) = \frac{1}{2}
(\frac{4\pi G}{3} \overline{\rho_H}(R_S)-\Omega^2)\\
\gamma & = & \frac{\Phi'_H(R_S)}{2R_S} = 
\frac{2\pi G}{3} \overline{\rho_H}(R_S)\nonumber
\end{eqnarray}

with $'\equiv d/dR$ and 
$\overline{\rho_H}(R)$  the mean halo density in a sphere of
radius $R$.
The inertial force of rotation and the Coriolis force cannot be put as the
gradient of a scalar potential. Eq. (\ref{phiterm}) 
tells us that the tidal potential 
$\Phi^{\rm tidal}(\mbox{\boldmath $r$})$ 
can be written as a contribution,
$\Phi^{\rm tidal}_{\mbox{\boldmath $r$}}(r) \equiv \beta r^2$,
that gives rise to an isotropic radial force, $\mbox{\boldmath
$F$}^{\rm tidal}_{\mbox{\boldmath $r$}}$,
and contributions giving rise to 
forces in the direction of $\mbox{\boldmath $R$}_S$ and $\mbox{\boldmath
$\Omega$}$.
Taking in $S_S$ a
cartesian coordinate system with the $X$ axis  pointing towards
$\mbox{\boldmath $R$}_S$,
$\mbox{\boldmath $e$}_x= \mbox{\boldmath $R$}_S/R_S$, the $Z$ axis  defined by
$\mbox{\boldmath $e$}_z= \mbox{\boldmath $\Omega$}/\Omega $
and $\mbox{\boldmath $e$}_y$ such that $\mbox{\boldmath $e$}_z \times
\mbox{\boldmath $e$}_x = \mbox{\boldmath $e$}_y$,
the tidal potential can be written as a quadratic form:

\begin{equation}
\Phi^{\rm tidal}(x,y,z)= \alpha x^2 + \beta y^2 + \gamma z^2 ,
\label{phitidal}
\end{equation}

giving rise to forces in the three orthogonal directions
(hereafter, $\mbox{\boldmath $F$}^{\rm tidal}_{\mbox{\boldmath $R$}}$, 
$\mbox{\boldmath $F$}^{\rm tidal}_{\mbox{\boldmath $\Omega$} \times \mbox{\boldmath
$R$}}$
and $\mbox{\boldmath $F$}^{\rm tidal}_{\mbox{\boldmath $\Omega$}}$, respectively).
In Figure \ref{fig1} we 
represent the intensity of the three components of the tidal 
force, for one of the models studied in this paper (see \S 3 and Table 3), as
function of ${\boldmath R}_S$.
If the satellite is in circular motion then $\Omega^2= \Phi'_H(R_S)/R_S$, and
$\Phi^{\rm tidal}_{\mbox{\boldmath $r$}}(r)=0$.  
In the general case, $\beta < 0$ ($\beta > 0$) at the pericenter (apocenter)
of the satellite orbit, while $\alpha < 0$ and $\gamma > 0$ anywhere
in the orbit. So, the $\mbox{\boldmath $F$}^{\rm tidal}_{\mbox{\boldmath $R$}}$
($\mbox{\boldmath $F$}^{\rm tidal}_{\mbox{\boldmath $\Omega$}}$) force changes its 
intensity being always disruptive (compressive),
while the radial tidal force changes its sign and intensity
as the satellite travels, being maximally disruptive at the pericenter
and maximally compressive at the apocenter. 

Defining  an effective potential 
$\Phi_{\mbox{\em eff}}(\mbox{\boldmath $r$};\mbox{\boldmath
$R$}_S,\mbox{\boldmath $\Omega$})$ as the total 
potential felt by the P particle, 

\begin{equation}
\Phi_{\mbox{\em eff}}(\mbox{\boldmath $r$};\mbox{\boldmath
$R$}_S,\mbox{\boldmath $\Omega$})= 
\Phi_S(\mbox{\boldmath $r$}) + \Phi^{\rm tidal}(\mbox{\boldmath
$r$};\mbox{\boldmath $R$}_S,\mbox{\boldmath $\Omega$}),
\label{potencialefectivo}
\end{equation}

then the energy (Jacobi integral)

\begin{equation}
E_J= \frac{1}{2} \left(\frac{d\mbox{\boldmath $r$}}{dt}\right)^2 +
\Phi_{\mbox{\em eff}}(\mbox{\boldmath $r$};\mbox{\boldmath
$R$}_S,\mbox{\boldmath $\Omega$})
\end{equation}

is not in general conserved

\begin{equation}
\frac{d E_J}{dt} = - \mbox{\boldmath $\dot{r}$} \cdot (\mbox{\boldmath
$\dot{\Omega}$} \times \mbox{\boldmath $r$}) +
\frac{\partial \Phi_{\mbox{\em eff}}}{\partial t}.
\label{varEJ}
\end{equation}

Note that in the reference system $S_S$, 
$\partial \Phi_{\mbox{\em eff}}/\partial t
\neq 0$  because $\dot{R_S}(t) \neq 0$. Then, if the satellite is in 
circular motion, $\dot{R_S}(t)=0, \mbox{\boldmath $\dot{\Omega}$} = 0$ and 
$E_J$ is conserved along the trajectory
of the P particle.

The angular momentum of the P particle is in general not conserved
in $S_S$

\begin{equation}
\frac{d \mbox{\boldmath $L$}}{dt}  =  
 - \mbox{\boldmath $r$} \times (\mbox{\boldmath $\dot{\Omega}$} \times
\mbox{\boldmath $r$}) - 2 \mbox{\boldmath $r$} \times
(\mbox{\boldmath $\Omega$} \times \mbox{\boldmath $\dot{r}$}) +
2\frac{\beta-\alpha}{R_S^2}
(\mbox{\boldmath $r$} \cdot \mbox{\boldmath $R$}_S)
\mbox{\boldmath $r$} \times \mbox{\boldmath $R$}_S + 
 2\frac{\beta-\gamma}{\Omega^2} (\mbox{\boldmath $r$} \cdot \mbox{\boldmath
$\Omega$}) \mbox{\boldmath $r$}
\times \mbox{\boldmath $\Omega$},
\label{varL}
\end{equation}

except for particles that move in radial trajectories along 
 the $\mbox{\boldmath $\Omega$}$ axis.

\section{Halos, Orbits and Galaxies}
\label{proforb}

As an external field we have taken the potential due to a massive halo, with
a mass distribution corresponding to a density profile given  by:

\begin{equation}
{\rho_{H, N}(R) \over \rho_{crit}} = {\delta_{c} \over (R/R_{C})(1 +
R/R_{C})^{2}}.
\label{Navprof}
\end{equation}
with $\rho_{crit}$ the critical energy density corresponding
to a flat geometry.
This is an accurate analytical fit over two decades in radius
and four orders of magnitude in mass
to the equilibrium density profiles of dark
matter halos which form in high resolution N-body simulations
in hierarchically clustering universes (Navarro, Frenk \& White  1996). They
are
characterized by two parameters: a scale radius, $R_{C}$, and a characteristic
dimensionless density, $\delta_{c}$, which in turn are correlated.
Note that $\overline{\rho}_{H, N}(R)/\rho_{H, N}(R) $ takes values
in the interval (1.50 , 3.17) for $0 \leq \frac{R}{R_C} \leq 2.5$.
This form of the profile has been chosen because we intend to
analytically describe halos with different masses,
so that different physical situations in which tidal forces play
an important role can be globally considered. Specifically, in this work we
will study orbits inside halos typically corresponding to
galaxy clusters and galaxy compact groups (c and g halos, respectively).
In Table 1 we give
the particular parameter values we have used.
This corresponds to a mass inside the virial radius, $R_{200}$
(the radius inside
which the overdensity is 200), of $M_{200} = 1.74 \times 10^{15}$
M$_{\odot}$ and $M_{200} = 2.1 \times 10^{13}$ M$_{\odot}$, that is,
about the typical
mass of a galaxy cluster and a compact group, respectively.
Note that the tidal field (and its radial gradient) produced by a compact
group like halo, at its characteristic length ($R_C=40$ kpc),
is stronger than that produced by a galaxy cluster like halo, at its typical
$R_C=600$ kpc.
Dark matter halos
have been represented analytically
by a continuous function because we are mainly interested in
exploring the possibility that tidal quasi-equilibrium
configurations are realized in Nature, and not in
studying the effects of
the dynamical friction force between the galaxy and the
 halo, that the fluctuating forces arising from their discrete
character would cause. 
Otherwise, tidal and dynamical friction effects
could not have been properly disentangled,
 as it is often the case in the literature.

The definition of the tidal radius is more sound from a physical point
of view when the satellite is in uniform circular motion
(see $\S$~4).
So, as a first test, satellites have been put on circular orbits
for  halos corresponding to galaxy clusters (c model).
Small $R_S$ values have been selected because
the tidal effects are stronger in the central regions of the
halo.
Also, general orbits
have been considered, in this case for compact group-like halos (g model).
Parameters characterizing these orbits are listed
in Table 2.

Concerning galaxies,
numerical simulations of the gravitational collapse in a cosmological
framework show that collapsed bodies are spherical symmetric in their
relaxed central zones. So, as a simplifying hypothesis, we will assume that
the satellite galaxy is spherically symmetric and has an isotropic velocity
dispersion tensor.
Note, however, that these symmetries will
only approximately hold for a satellite
particles that move according with Eq. (\ref{eqorfin}), particularly those
whose apocenters lie in the outskirts of the configuration.

Initially, self-gravitating spherically symmetric  configurations
will be taken to be t-limited King spheres
with an isotropic velocity dispersion tensor, i.e., King spheres
with their tidal radius determined by the tidal field.
 They are based on the so-called
  King-Michie distribution function (DF), $f(r,E)$, that is an
approximative stationary
solution of the Boltzmann equation with a Fokker-Planck collision
term (King 1965, 1966; Michie 1963). These DF are lowered Maxwellians,
with a cut-off at the escape velocity to the border of the
configuration for the less bound particles at each position.
This escape velocity can be written as:

\begin{equation}
v^2_{esc}(r)=2 \left( K -  \Phi(r) \right)
\label{vesc}
\end{equation}

where $\Phi(r, R_S)$ is the total potential felt by the satellite particles
and $K$ is a constant defining the zero point of the potential.
In terms of the shifted energy, $\varepsilon(r) = \Phi(r,  R_S) + v^2/2 - K$,
the King-Michie DF is zero for $\varepsilon(r) > 0$ and for
$\varepsilon(r) \leq  0$ it is given by:

\begin{equation}
f(r,v)= k \exp[W(r)-W_0] [\exp(-j^2 v^2)-\exp(-j^2 v^2_{esc}(r))]
\label{fKing}
\end{equation}

where      $j^2=1/2\sigma^2_0$; $\sigma_0$ is  an approximation to
the 1-dimensional
velocity dispersion at the center of the configuration;
$W(r)=2j^2 \left(K - \Phi(r, R_S) \right)$ is the dimensionless potential;
$W_0 \equiv W(0)$ is a parameter of the model
and $k$ is a normalization constant.
Standard King-Michie spheres take  $\Phi(r,  R_S) = \Phi_S(r)$
and they have $r_{\rm t}$ as a free parameter.
Other free parameters for these isolated spheres are the dimensionless
	central potential, $W_0$, the 
	approximate central velocity dispersion, $\sigma_0$,
the core radius, $r_o$, and the total mass, $M_S$. Note that only
three of them are independent (e.g. Binney \& Tremaine 1987).

In an external field, $r_{\rm t}$ is determined by the external forces
and only two more parameters for t-limited King models
are left free.
To our knowledge, no description of the method to build-up King spheres with
a prefixed $r_{\rm t}$ can be found in the literature.
So, we will briefly comment on it.
First, we note that 
inside a spherically symmetric satellite we must have an
{\it isotropic} potential. However, the tidal potential
is not spherically symmetric, therefore
we need to approximate the inner tidal potential field,
$\Phi^{\rm tidal}({\boldmath r}; R_S)$ (Eq. (\ref{phitidal}), 
to an isotropic field, $\Phi_{\rm radial}^{\rm tidal}(r; R_S)$
(see $\S$\ref{movcircular} for a discussion on the approximation).    
Now, to build-up these t-limited King spheres,
the Poisson equation for the satellite potential, $\Phi_S(r,  R_S)$,
must be solved with a density given by the  King-Michie  DF
(Eq.~(\ref{fKing})),
which, on its turn, depends on the total potential
$\Phi(r, R_S) =
\Phi_S(r,  R_S)  +  \Phi_{\rm radial}^{\rm tidal}(r; R_S)$            

\begin{equation}
\rho_S(r) = \int d {\bf v} f(r,E) =
 {\rho_0 \over \Gamma(5/2, W_0)} {\rm exp} \left[ W(r) - W_0 \right]
 \Gamma(5/2, W(r))
\label{rhoII}
\end{equation}

where  $\rho_0 = \rho_S(0)$
and $\Gamma(\alpha, W)$ is the incomplete
gamma function.

In order to solve the Poisson equation, appropriated boundary conditions have
to be specified. First, as usual,
the net force at the center of the configuration
must vanish
$\left( {d W(x) \over dx} \right)_{x=0} = 0$ and $W(0) = W_0$.
Moreover,
 given a satellite of
mass $M_S$, the tidal field fixes
its tidal radius and one must have
$M(x_{\rm t}) = M_S$,  
(or equivalently
$W(x_{\rm t}) = 0$), where $x_{\rm t} = r_{\rm t}/ r_o$ and
\begin{equation}
M(x_{\rm t}) = 4 \pi r_o^{3}
\int_0^{x_{\rm t}} x^2 dx \rho_S(W(x)).
\label{masa}
\end{equation}

To build-up t-limited King spheres in a given point, $R_S$, of the satellite
orbit, characterized by $E_H$ and $ L_H$,
the following practical procedure has been used:
i) we choose as free parameters of the configuration $M_S$ and
$r_o$, ii) the $W_0$ parameter is determined by the condition
$W(x_{\rm t}) = 0$ and iii) Eq (\ref{masa}) and $M(x_{\rm t}) = M_S$
gives the central density, $\rho_0$, and then the relation (King 1966)
\begin{equation}
{4 \pi G r_o^{2} \rho_o \over \sigma_0^{2}} = 9
\label{relking}
\end{equation}
gives $j^2=1/2\sigma^2_0$, that is, the $\sigma_0$ parameter.

Once an orbit has been selected, the t-limited King models
corresponding to the tidal field at different points in the orbit
have been obtained.
The values of the satellite mass, $M_S$, and core radius, $r_o$,
used as input are $M_S = 2.2 \times 10^{11} M_{\odot}$,
$r_o = 2.4$ kpc for galaxy models on c-type orbits,
and $M_S = 1.3 \times 10^{11} M_{\odot}$, $r_o = 1.2$
kpc for galaxy models on g-type orbits.
Different models for the tidal radius have been considered
(see $\S$\ref{results}).
In Table 3 we  give these tidal radii and in Table 4 we give their 
corresponding $W_0$ and $\sigma_0$ values. Note that
 both $\sigma_0$ and $M_S$ are within the
  observationally allowed ranges for typical elliptical galaxies.

Once the velocity DF (given by Eq. (\ref{fKing}) with the corresponding
parameters) and the density profile
(given by Eq. (\ref{rhoII})) have been determined, a galaxy represented
by a Montecarlo  realization  of these velocity DF and density profile,
with 10000 particles, has been built up. Galaxies are non-rotating in the
$S_S$ rotating frame and so the tidal potential is time-independent
as far as $R_S$ is constant along the orbit.
 These systems have been left to evolve during a time
interval of 12.5 Gyears, under the gravitational forces caused by both,
the particle configuration and the dark matter halo.
A treecode algorithm (Hernquist 1987), modified to take into account
the external force caused by the density 
distribution given by Eq. (\ref{Navprof})
acting on each satellite particle, P,
has been used to integrate the motion equations.
Note that the approximations discussed in $\S$2 are
 not used at this stage;
 these approximations are only
used to build up the {\it initial} configurations of the
self-gravitating satellites (see $\S$\ref{results}).
The neglect of the stochastic forces caused by
the discrete character of the halo is not likely to substantially 
modify the results we obtain. 
The study of the dynamical friction effects must be made for each 
particular case, however,
no important effects can be expected both for a cluster like halo
of $M_{200} = 1.74 \times 10^{15}$ (see Klypin et al. 1999, their
Figure 7), or a compact group like halo when the velocity
dispersion of the orbiting galaxies is of the order of that of
the halo itself (see Tables 2 and 4, and Eq. (53) of 
Dom\'{\i}nguez-Tenreiro \& G\'omez-Flechoso 1998, where it is shown
that, when this is the case, the dynamical friction timescales
can be considerably longer than those predicted by the
popular Chandrasekhar 1943 formula). 

\section{Results}
\label{results}

\subsection{Satellite in uniform circular motion}
\label{movcircular}

In this case $R_{\rm S} $ and  $\Omega $ do not change, and
$\alpha =  2 \pi G ( \rho_H(R_S) -
\overline{\rho_H}(R_S)), \beta = 0$ and
$\gamma = 2 \pi G \overline{\rho_H}(R_S)/3$.
$E_J$, and, consequently, $\varepsilon_J = E_J -  K$, are integrals of the
motion for the satellite particles as it orbits inside the halo.
The gradient of the tidal potential, 
$ \nabla_{r} \Phi^{\rm tidal}({\bf r};{\bf R}_S,{\bf \Omega})$,
makes a contribution to 
$\nabla_{r}  \Phi_{\mbox{\em eff}}({\bf r};{\bf R}_S,{\bf \Omega})$
that has the same sign as $\nabla_{r} \Phi_S(r)$ 
along the  $z$ direction,
while it has the opposite sign along the $x$ direction
 (see Eqs.  (\ref{phitidal})).
 The effective potential has a saddle  point at positions
 $L_{X}^{\pm} = (\pm x_{\rm e}, 0, 0)$
  where

\begin{equation}
x_{\rm e}^3 = - \frac{G M_{\rm S}}{2 \alpha(R_{\rm S})},\hspace{2cm}
\label{XE}
\end{equation}

  These points  are  Lagrange points,
  where the net force on a satellite particle
  vanishes, when terms of second order in $r/R_S$ or higher
are neglected, as in Eq. (5).
  The equipotential surface
  $\Phi_{\mbox{\em eff}}({\bf r};{\bf R}_S,{\bf \Omega}) =
  \Phi_{\mbox{\em eff}}(( x_{\rm e}, 0, 0); {\bf R}_S,{\bf \Omega})$
  is the corresponding Roche surface; this is the last closed zero velocity
  surface relative to the conserved Jacobi integral $E_J$.

As pointed out in $\S$1, the limiting radius, $r_{\rm t}$,
is usually defined in terms of the $x_{\rm e}$ distance: while King (1962)
takes $r_{\rm t, K} = x_{\rm e}$, Keenan (1981b) proposes 
$r_{\rm t, Kee} = 2 x_{\rm e}/ 3 $.
As a first step to study {\it quasi-equilibrium} configurations, 
we have built-up t-limited King spheres, with limiting radii equal
to both, $r_{\rm t, K}$ and $r_{\rm t, Kee}$, and moving along a circular orbit
(with parameters as specified in Table 2) inside a cluster-like
halo, characterized by parameters as specified in Table 1.
The zero point of the potential for these spheres has been taken to be:

\begin{equation}
K(R_S) = \Phi_{\mbox{\em eff}}((x_{\rm e}, 0, 0); R_S,{\bf \Omega}) 
 = - \frac{3 G M_{\rm S}}{2 x_{\rm e}(R_S)}
\label{zeropoint}
\end{equation}

In this way we ensure that the zero velocity surface for 
$\varepsilon_{\rm J} = 0$
particles (i.e., the less bound ones) is also the limiting surface of
the configuration (ideally defined as the equipotential surface where
the less bound particles and with minimum angular momentum have
their apocenters, see G\'omez-Flechoso
1997 for a discussion).

To test out that the results on the evolution of the galaxy models
presented in this paper are free from two-body effects, the evolution
of {\it isolated} King models, corresponding to the
galaxy models in Tables 3 and 4,  has been followed in control
simulations. After 12.5 Gyears, 
no two-body effects have been detected in any case at
a significant level. As an illustration,
in Figures \ref{fig2} we plot the evolution of the radii
enclosing a 75\% and a 95\% of the isolated satellite
 initial mass,  normalized to the corresponding
limiting radii and referred to the center of density of the galaxy, for
one c-model (Figures \ref{fig2}a and \ref{fig2}b) and one g-model 
(Figures \ref{fig2}e and \ref{fig2}f). 

Let us now turn to the behavior of the galaxy models
orbiting on circular orbits inside the halo.
In Figures \ref{fig2}a and \ref{fig2}b we show the evolution of the radii
enclosing a
75\% and a 95\% of the satellite total initial mass, $M_S$,
for c-Kee and c-K galaxy models, normalized to the corresponding
tidal radii. These radii are referred to the center of density of the galaxy.
In these Figures, we see that, for both prescriptions, 
the increase of these radii 
 is nearly imperceptible, and in fact, 
the comparison with the behavior of the isolated King model
indicates that
the almost negligible 
rate of orbit expansion can be entirely accounted
for as a result of two-body heating. 

Another way to quantify the diffusion in position space is to analyze
the evolution of $M(r,t)/M_S$,
the mass inside radius $r$, normalized to the 
initial satellite mass, $M_S$. This is plot in Figure \ref{fig3} for the c-Kee
model at $t =$ 0, 6.25 and 12.5 Gyrs.
This Figure indicates that at $t$= 6.25 Gyrs, 
only the 0.8\% of the initially bound particles are beyond the tidal radius
(similar result is obtained for c-K model).
At the end of the simulation, 
the 2.3\% of the  particles are at $r > r_{\rm t}$. 
However, these minor changes result from two-body evolution, and, moreover, 
in any of c models, the innermost  volumes of the galaxies 
(inside a sphere of, say, $r/r_{\rm t} \simeq 0.3$) are absolutely 
not affected
by the evolution.  

Let us now  analyze the evolution of the particle
velocity distribution. The average velocity dispersion
of particles that remain at the configuration does not
appreciably change.   
In Figure \ref{fig4}a 
we represent $M(>v,t)/M_S(t)$ ,i.e., the fraction of particles
hotter than a given $v$,
versus $v$ normalized to the initial 3-D velocity dispersion 
at several times for the c-Kee prescription
(escapers have not been included).
No evolution
is detected at a significant level for this model.
The behavior of the c-K model is similar.

The anisotropy parameter 
for a given particle sample, 
$\beta_{\rm an}  = 1 - \sigma^2_{\theta}/\sigma^2_r$,
where  $\sigma_{\theta}$ ($\sigma_r$) is the 
tangential (radial) velocity dispersion 
 for the sample in consideration,
quantifies how much a given velocity distribution deviates from one
with an isotropic velocity dispersion tensor, that would have
$\beta_{\rm an} = 0$.
In Figure \ref{fig5} we show the evolution of the  
 $\beta_{\rm an}  $ parameter for configuration particles outside the
radius enclosing the 67\% of the bound (i.e., excluding escapers) mass
at each time.
Both models exhibit similar amounts of anisotropy (roughly $\simeq 0$),
indicating that the evolution of the external orbits (the most
affected by the tidal field) is negligible.

The negligible amount of change along 12.5 Gyears of evolution indicates 
that both the c-K and c-Kee models are quasi-equilibrium configurations
in the cluster tidal field. As explained above,
 $r_{\rm t,K}$ and $r_{\rm t, Kee}$ correspond to the semiaxes
of the galaxy Roche surface along the OX and OY directions,
respectively.
 To gain some insight into the physical
basis of this quasi-stable
configuration, let us recall that for a spherically symmetric system, 
the escape velocity to the border of the configuration
at position $r$ can be written:

\begin{equation}
v_{esc}(r; R_S)= ( 2 | K(R_S) - \Phi_S(r) - \Phi^{\rm tidal}_{\rm radial}(r;
R_S) |)^ {1/2}
\label{VescTidal}
\end{equation}

once the isotropic approximation $\Phi_{\rm radial}^{\rm tidal}(r; R_S)$
 for the tidal potential is considered. 
A stable configuration needs to have zero escape velocity
at its limiting radius, $v_{esc}(r_{\rm t}; R_S) = 0$.
This condition ensures the stability of the configuration
because the inner escape velocity field of the King model satisfies the
boundary conditions imposed by the tidal field,
so that the effective potential vanishes at $r_{\rm t}$. 
But taking $r_{\rm t} = r_{\rm t,K} = x_{\rm e}$,
the $v_{esc}(r_{\rm t}; R_S) = 0$ condition 
and Eq. (\ref{VescTidal}) demand
that at $r_{\rm t}$ the
 tidal potential is
$\Phi_{\rm radial}^{\rm tidal}(r_{\rm t}; R_S) = 
\alpha(R_S) r_{\rm t}^{2} = \Phi_{\rm radial, \alpha}^{\rm tidal}(r_{\rm t}; R_S)$, 
with $\alpha(R_S)$ given by Eq. (\ref{XE});
taking $r_{\rm t} = r_{\rm t,Kee} = 2x_{\rm e}/3$
the  tidal potential at $r_{\rm t}$ must be
$\Phi_{\rm radial}^{\rm tidal}(r_{\rm t}; R_S) =
\beta(R_S) r_{\rm t}^{2}=\Phi_{\rm radial,\beta}^{\rm tidal}(r_{\rm t}; R_S)$ 
(= 0 for circular orbits).
By continuity requirements, 
the limiting radii prescription in the c-K and c-Kee models
can thus now be looked at as a result of two different choices
of the $\Phi_{\rm radial}^{\rm tidal}(r; R_S) $
potential  field 
under the condition that $v_{esc}(r_{\rm t}; R_S) = 0$,
namely $\Phi_{\rm radial, \alpha }^{\rm tidal}(r; R_S) =
\alpha(R_S) r^{2}$ and 
$\Phi_{\rm radial, \beta }^{\rm tidal}(r; R_S) =
\beta(R_S) r^{2}$.
A third possible choice for the
$\Phi_{\rm radial}^{\rm tidal}(r; R_S) $ potential is
to take 
$\Phi_{\rm radial, \gamma}^{\rm tidal}(r; R_S) = \gamma(R_S) r^2$
that results, under the condition of zero escape
velocity at $r_{\rm t}$, in a limiting radius,
$r_{\rm t, \gamma}$, corresponding to the semiaxis
of the galaxy Roche surface along the OZ direction.
Hereafter, the limiting radii satisfying the zero escape
velocity condition for
a given choice of $\Phi^{\rm tidal}_{\rm radial}(r; R_S)$
will be termed after this choice. In  Table 3 (first row)
we give the corresponding numerical values for models of 
the galaxy under consideration, and of the 
halo and circular orbit as specified in Tables 1 and 2.
Note that $r_{\rm t, \alpha}$ (or $r_{\rm t,K})
> r_{\rm t, \beta}$ (or $r_{\rm t, Kee}) > r_{\rm t, \gamma}$.
For completeness, in Table 3 we also give the limiting
radii, $r_{\rm t, M}$, corresponding to the case where
$\Phi^{\rm tidal}_{\rm radial}(r; R_S)$  is taken to be
the  monopolar
component of the  tidal field potential expansion 
(Eq. (6)) into spherical harmonics, namely:
 
\begin{equation}
\Phi_{\rm radial, M}^{\rm tidal}(r; R_S) =
(\alpha(R_S) + \beta(R_S) + \gamma(R_S)) r^2/3.
\label{MonoComp}
\end{equation}                            

The tidal radius $r_{\rm t, M}$ represents 
a kind of {\it average radius} of the 
Roche surface. 
The results described above concerning the near-constancy of the 
configuration along 12.5 Gyears, 
both for the
$r_{\rm t, K}$ and $r_{\rm t, Kee}$
choices,
with mass losses compatible with that expected from two-body
heating, suggests that the ambiguity in the
 precise
prescription for $r_{\rm t}$, has no  consequences, with the different
choices being essentially equally good, provided that the initial
configuration is a quasi-equilibrium one.
This non-evolving character of quasi-equilibrium
collisionless self-gravitating configurations as they move
on circular orbits inside a tidal field, can be easily
understood in the framework of the adiabatic protection
hypothesis (see $\S$\ref{GenMotion}), because the time-independent 
character of the tidal field in the rotating $S_S$ frame
would imply zero escape rates.

\subsection{Satellite in General Motion}
\label{GenMotion}

If the satellite is not in circular motion, the intensity of the 
tidal forces changes as the satellite orbits,
being maximum (minimum) at pericenter (apocenter) passage 
(see Figure \ref{fig1} for an illustration).
Moreover, in the $S_S$ frame, a force term in
${\bf F_{ \dot{\Omega}}} = -{\bf \dot{\Omega}} \times {\bf r}$
 appears that cannot be 
expressed as the gradient of a potential; however, 
the disruptive effects of the tidal forces are more important than
those of this term for almost all the satellite particles along
the orbit\footnote{The heating due to ${\bf F_{ \dot{\Omega}}}$ can only
produce the loss of a reduce number of particles that have a 
large internal angular momentum parallel to the angular
momentum of the satellite orbit},
and particularly so at pericenter where this
term vanishes, so that 
it can be safely neglected in this work
in what concerns the set-up of the initial configurations
(this term is not neglected in the numerical models, where, as
explained above, the exact force produced by the halo on each
satellite constituent particle has been considered).

Because $\Phi_{\em eff}(R_S)$ is not constant, neither $E_J$ nor
$\varepsilon_J$ are conserved, as deduced from Eq. (11).
The effective potential $\Phi_{\em eff}$ changes with a
timescale given by the anomalistic period of the satellite, $T$,  
(i.e., the time interval between two successive pericenter passages).

When a particle of energy $E$ orbits inside a  time-dependent potential
with sideral period $\tau$, its energy can be increased or decreased as
time passes; in most cases, the average energy of a particle system
in a time-dependent potential tends to increase (see Spitzer 1987).
The relative energy change in a given orbit per revolution can be 
very small if $T/\tau > 1$. In fact, it has been shown (Kruskal 1962)
that it goes to zero faster than any power of $T/\tau$, as, for example,
does an exponential function of $- A T/\tau$ (with  $A$ a dimensionless
constant of order unity, see Spitzer 1987),  
 provided  that the
$T$ and $\tau$ periods are not commensurate quantities, otherwise resonance
phenomena can occur (Weinberg 1994).
So, an energy gain occurs when $T < \tau$ and it 
would result in an orbit expansion. If this happens
for an important fraction of the satellite constituent particle orbits,
the satellite will expand and loss mass 
at a rate similar to the expansion rate. On the contrary,
if $T > \tau$ for most of the constituent particles (or $T > t_{\rm dyn}$,
where $t_{\rm dyn}$ is a dynamical time measuring an average period
for the satellite constituent particles),   then only moderate
satellite heating can be expected: the system is said to be adiabatically
protected. 
In any case, as mass loss rates are similar
to the satellite expansion rate due to tidal
heating, it can be expected that the relative mass losses of a galaxy with
a dynamical time scale $t_{\rm dyn}$ orbiting with a period $T$
go to zero approximatively as $\exp( - A T/t_{\rm dyn})$.

Now, let us point out that only when mass losses are unimportant,
that is, when the system is adiabatically protected, is {\it quasi-equilibrium}
in a tidal field a physically sound concept: tidal quasi-equilibrium
demands low mass losses as the satellite orbits.
But in this case the energies of most satellite particles
will be conserved to a good approximation, and, then, the equilibrium
tidal radius can be calculated following the same  physical reasoning
as discussed in $\S$\ref{movcircular}, namely $r_{\rm t}(R_S)$
must be taken to be the radii of the Roche surface along
the three axis, or, equivalently, must be taken 
such that
the escape velocity field of the King model satisfies the conditions
imposed by the external tidal field 
($v_{esc}(r_{\rm t}(R_S); R_S) = 0$, with $v_{esc}(r; R_S)$ 
given by Eq. (\ref{VescTidal})
 and with the different choices for the isotropic component
of the tidal field as discussed  in $\S$\ref{movcircular}). 
Moreover, in this case, a new ambiguity appears concerning the 
matching procedure, as the
external potentials, and, consequently, $x_{\rm e}$, 
depend on the satellite orbital phase $R_S$.
The most natural choice is to take $R_S^{eq} = R_p$
($R_S^{eq}$ is the point of the orbit where the initial 
configuration  is built up), 
because the system suffers a kick at pericenter passage.
In this work other possibilities have also been
considered to quantify how much the system evolution depends
on the orbital point where the initial equilibrium configuration
is built up: $R_S^{eq} = R_a$ (apocenter distance)
 and $R_S^{eq} = R_0$
(radius of the circular orbit with the same energy $E_H$ as the
eccentric orbit under consideration). Table 3 summarizes the
different possibilities
we have considered and gives the 
corresponding limiting 
radii $r_{\rm t}$.

Note that several models in this Table have similar $r_{\rm t}$
values:
a) those with $r_{\rm t}(R_S^{eq}) \simeq r_{\rm t, \beta}(R_p)$
(g-p-$\beta$,  g-p-$\gamma$  and g-p-M models), b) those with
$r_{\rm t}(R_S^{eq}) \simeq 1.5 r_{\rm t, \beta}(R_p)$
(g-p-$\alpha$, g-$R_0$-$\beta$,  g-$R_0$-$\gamma$  and g-$R_0$-M models),
c) those with
$r_{\rm t}(R_S^{eq}) \simeq 2 r_{\rm t, \beta}(R_p)$
(g-$R_0$-$\alpha$, g-a-$\beta$,  g-a-$\gamma$  and g-a-M models),
and finally, d) the g-a-$\alpha$ galaxy model whose
 $r_{\rm t}(R_S^{eq})$
 is about $3.5 r_{\rm t, \beta}(R_p)$.
 In Table 4 we give other galaxy parameters of interest 
 (see $\S$3) for the models in Table 3.
In particular, an average dynamical timescale $t_{\rm dyn} = (3 \pi/16 G
\overline{\rho})^{1/2}$ (Binney \& Tremaine 1987)
for these models is given in this Table.

These different models have been left to evolve as they orbit inside
a compact group-like halo (see Table 1) following an eccentric g-like
orbit (see Table 2) during 12.5 Gyears.
Accordingly with the discussion above, it can be expected that
the relative mass losses for these different models at any given time
be  approximatively 
proportional to $\exp(-A T/ t_{\rm dyn})$ if adiabatic protection 
is at work in these simulations.
To show that this is the case, 
in Figure \ref{fig6} 
 we have plot  $T/t_{\rm dyn}$ versus the logarithm of $\Delta M_S/M_S$ 
after 12.5 Gyears
of evolution for the g-model galaxies  in Table 3, where $\Delta M_S$  is
the total mass lost in escapers  at 12.5 Gyears and $M_S$ is the initial
satellite mass.
A very good linear relation appears.
It corroborates the adiabatic
protection hypothesis, that for a galaxy on circular motion
predicts zero mass losses (because $T = \infty$), as we have
found in  $\S$\ref{movcircular}. 

Let us now describe different quantitative aspects of the satellite
evolution. In Figure \ref{fig7} we plot the  $M_S(t)/M_S$ evolution
for six g-models in Table 3 ($M_S(t)$ is the bound mass at time $t$,
defined as the mass of the system whose total internal energy $E<0$).
Note that mass losses are quite linear and regular. No evidences
of the perigalactic passages can be seen because of the small amount of 
mass losses at each passage.
This Figure is an illustration that 
g-models in Table 3 exhibit different degrees of mass losses
accordingly with the classification above:
a) those with $r_{\rm t}(R_S^{eq}) \simeq r_{\rm t, \beta}(R_p)$
that
loss very few mass in 12.5 Gyrs of evolution  ($\simeq$ 1.0\%);
b) those with $r_{\rm t}(R_S^{eq}) \simeq 1.5 r_{\rm t, 
\beta}(R_p)$, 
that have up to a 5.8\% of escapers;
c) those with
$r_{\rm t}(R_S^{eq}) \simeq 2 r_{\rm t, \beta}(R_p)$,
where the mass losses are not negligible (g-$R_0$-$\alpha$ and g-a-$\beta$
galaxy
models, with a 12.8\% and 14.6\% of escapers at 12.5 Gyrs);
and, finally, d) the g-a-$\alpha$ galaxy model, where mass losses are
more significant (33.5\%).

 Figures \ref{fig2}  for the g-p-$\beta$, g-p-$\alpha$, g-a-$\beta$ and 
g-a-$\alpha$ galaxy models (c, d, e and f),
indicate that the mass loss is maximum for the g-a-$\alpha$  model,
where the shell corresponding to the 25\% outsiders is lost
at $t  \simeq 7 $ Gyrs, and minimum for the g-p-$\beta$ galaxy,
where the radius enclosing the 95\% inner particles had not yet crossed
$ r_{\rm t, \beta}$ at $t  \simeq 12.5 $ Gyrs.
The g-p-$\alpha$ and g-a-$\beta$ models show an intermediate behavior.
Other models, not shown in the Figure \ref{fig2} (see Table 3),
exhibit different evolutionary trends according with the classification above.

These  trends are confirmed by Figure \ref{fig3}, 
where we can see that  
the inner regions of the configuration 
(say, $r/r_{\rm t} < 0.2$) are not very much affected by the
tidal forces for g-p-$\alpha$ model.

The analysis in the velocity space indicates that 
$\sigma$ for particles that remain at the configuration 
does not change significantly; also, the percentage of  particles
in each velocity bin remains roughly  constant along the 12.5
Gyears of evolution.
All the g-models in Table 3 have a similar qualitative behavior,
but heating is maximum for the g-a-$\alpha$ model and minimum and negligible
for the g-p-$\beta$ model. An average behavior is exhibited by the 
g-p-$\alpha$ model (see Figure \ref{fig4}b).

The anisotropy evolution (Figure \ref{fig5}) confirms these findings and is,
for any g-model in Table 4, 
always indicative that particles on more elongated orbits
are more likely to escape (Keenan 1981a, 1981b).

The results so far described indicate that once a 
quasi-equilibrium galaxy model is
built-up at a given point of its orbit, the configuration
does not appreciably change as the galaxy orbits,
 except for mass losses.
The system does not relax towards the different equilibrium solutions
corresponding to the different points of its trajectory. 

\section{Summary and Conclusions}
\label{sumcon}

In this paper we address the issue of the existence of quasi-equilibrium  
self-gravitating configurations in a quiescent tidal field,
that is, the possibility that self-gravitating configurations
exist that are able to survive for a Hubble time or so in a given dense
environment without being tidally stripped or disrupted.
More specifically, we have tried to answer  the question of how
to build-up such configurations with their limiting radius 
determined by the tidal field and their remainder characteristics
described by  parameters that can be fixed a priori.

As the simplest hypotheses, the tidal field produced by a static,
spherically symmetric, dense, extended halo has been considered.
Also for simplicity, the configurations have been taken initially to be
spherically symmetric and to have an isotropic velocity tensor
(t-limited King spheres). They orbit inside the halos.
Both circular and eccentric orbits have been considered.
In both cases, quasi-equilibrium self-gravitating configurations
have been built-up by taking as tidal radii the radii of their
Roche surface along different axes, or, equivalently, by
defining the escape velocity field
of the configuration taking into account the
requirements imposed by the tidal field produced by the external halo.
The gravitational field inside the configuration is spherically
symmetric, while the tidal field has no this symmetry and one has to resort
to a choice of its radial component.
So, an ambiguity arises when matching the internal and external fields
of forces
at $r_{\rm t}$. Different possible choices have been considered.
In the case of an eccentric orbit the tidal field depends on the orbital phase
and a new ambiguity arises regarding the orbital position where the
matching is made. Here also different possibilities have been
explored.

To study the survival of the configurations, we have evolved Montecarlo
realizations of t-limited King 
galaxy models, orbiting in the tidal field produced
by a dark cluster-like halo (for circular orbits) or galaxy group-like (for
eccentric orbits) halo.
These galaxies have been taken to be collisionless
systems, i.e., such that heat transport and relaxation 
cannot proceed 
through two-body encounters,
but through the oscillations of their collective self-consistent
potential.

The general result of the simulations, irrespective on how the 
matching has been made, is that the bulk of the models are conserved 
along 12.5 Gyears of evolution
both for circular and eccentric orbits,
even if some mass losses occur in some cases.
A good linear relation between the ratio of the galaxy anomalistic
period to its dynamical timescale,
$T/t_{\rm dyn}$, and the logarithm of the
relative mass losses for the different galaxy models
has been obtained, suggesting that  adiabatic protection
 is at work in these simulations. In the case 
of galaxies on circular orbits, the adiabatic protection 
hypothesis predicts zero mass losses; the results of our
simulations for circular motion are  also in agreement with
this prediction, once the almost imperceptible 
two-body effects are considered.   
In the case of eccentric orbits,
if the galaxy configuration closely corresponds to the
tidal equilibrium solution at its actual environment,
 no important oscillations of the potential produced by
tidal forces can be expected, and, in fact, our simulations
show that 
 once the satellite
particles are distributed in positions and velocities 
according to the equilibrium solution at one given point
of the satellite orbit, they approximately remain so as
the satellite orbits, except for the marginally bound
particles, in which case most of them are lost to the
configuration.
Even if continuous mass losses are important in some cases,
the system does not relax towards the equilibrium solutions
corresponding to the different positions of the satellite,
presumably because these different equilibrium solutions
are not far enough to produce strong collective potential
oscillations. 
 
Configurations corresponding to equilibrium at pericenter, 
where the tidal forces are maximum, are those that suffer from 
less  escape (they are hyperstable solutions at the other
points of the orbit) and anisotropy development.
Among the simulations presented in this paper,
the maximum mass losses occur
  when the  component of the tidal force in the ${\bf R}_S$ direction
  is chosen as radial component of these forces
 and the initial configuration is prepared
 at the apocenter ($\simeq 33.5$\%
along 12.5 Gyrs). But, as stated above, it is mainly the
$T/t_{\rm dyn}$ ratio that determines the evolution rates
and mass losses.

The results described so far suggest that 
the configuration of spheroidal galaxies  is fixed at its 
formation, determined by its mass, energy content and the
environment at that moment. After formation, only 
moderate tidally induced evolution can be expected for a
galaxy living in environments  of density similar to that of its environment
at formation. 
These results  also suggest that a continuous slow mass
loss along long periods can occur, without   
destroying the system, if the density of the environment at
formation is lower than that of the environment at galaxy pericenter
passage.
The galaxy will be easily destroyed, however, should
it be placed at an external field whose 
corresponding equilibrium tidal radii is much 
smaller than the limiting radius of 
 the actual galaxy (see GD00 for a discussion).

\acknowledgments

It is a pleasure to thank l'Observatoire de Gen\`eve for its kind
hospitality while this work was finished.
M.A. G\'omez-Flechoso was supported by the Direcci\'on General de
Ense\~nanza Superior (DGES, Spain) through a
fellowship. It also supported in part this work, grants AEN93-0673,
PB93-0252 and PB96-0029.

\clearpage

{\bf FIGURE CAPTIONS}\\

\figcaption[gomezflechoso_fig1.ps]{The intensity, at $r = r_{\rm t}$, of the
components of the tidal force in the $\mbox{\boldmath $R$}_S, \mbox{\boldmath
$\Omega$} \times \mbox{\boldmath $R$}_S$
and $\mbox{\boldmath $\Omega$}$ directions,
normalized to the intensity of the $\mbox{\boldmath $F$}_S(r_{\rm t})$ force,
for the g-p-$\beta$ model (i.e., 
$r_{\rm t} = r_{\rm t, \beta}(R_p)$), as a function of
$R_S$. A negative value means that the tidal forces are {\it disruptive}.
For galaxy particles placed at distance $r$ from the galaxy center, the ratios 
of the intensities in the $\mbox{\boldmath $R$}_S$ and $\mbox{\boldmath
$\Omega$}$ directions
are obtained by multiplying the values in this Figure by
$(M_S/M_S(r))\left(r/r_{\rm t, \beta}(R_p) \right)^3$. \label{fig1}}

\figcaption[gomezflechoso_fig2.ps]{Evolution of 
the radius enclosing the 75\% (left panels) and the 95\% 
(right panels) of the galaxy total initial mass, $M_S$, 
normalized to the tidal radii corresponding to each model.
Several galaxy models, moving on either  circular (c)
or eccentric (g) orbits, are shown (see Tables 3 and 4),
and also galaxy models that evolve in isolation (iso).
\label{fig2}}

\figcaption[gomezflechoso_fig3.ps]{ The total galaxy mass inside radius $r$,
normalized to the galaxy total initial mass, $M_S$,
for $ t = 0, t = 6.25$ and $ t = 12.5$ Gyrs corresponding to 
c-Kee and g-p-$\alpha$ galaxy
models. Radii are given in units of their respective tidal radii  
(see Tables 3 and 4 and text).\label{fig3}}

\figcaption[gomezflechoso_fig4.ps]{The fraction 
of the galaxy total mass, with velocity higher
than $v$ at $t = 0$, $t = 6.25$ and $t = 12.5$ Gyrs, for 
 (a) c-Kee and (b) g-p-$\alpha$ galaxy models.
 Velocity is normalized to the 3-D initial velocity dispersion.
Escapers have not been taken into account.\label{fig4}}

\figcaption[gomezflechoso_fig5.ps]{The evolution of the 
anisotropy parameter $\beta_{\rm an}$ corresponding to the 33\% more
distant particles among those that
have not escaped at time $t$.
Results for the two c-type and 
several g-type models in Table 3 are shown.\label{fig5}}

\figcaption[gomezflechoso_fig6.ps]{The ratio of the anomalistic period
of the galaxy model, $T$, to its dynamical time, $t_{\rm dyn}$, versus the
logarithm of its relative mass loss in 12.5 Gyears for the 12
 g-type
models in Table 3.\label{fig6}}

\figcaption[gomezflechoso_fig7.ps]{Evolution 
of the fraction of the galaxy initial mass, $M_S$,
that remains at the configuration, for several g-model galaxies 
(see Tables 3 and 4).\label{fig7}}

\clearpage
\begin{deluxetable}{lcccc}
\tablecolumns{5}
\tablewidth{0pc}
\tablecaption{Halo parameters}
\tablehead{
\colhead{} & \colhead{$\delta_c$} & \colhead{$R_C$} & \colhead{$R_{200}$} &
\colhead{$M_{200}$}\\
\colhead{ }&  \colhead{ }         & \colhead{(kpc)} & \colhead{(kpc)}   &
\colhead{($10^{13}$ M$_{\odot}$)}}
\startdata
Compact Group  (g) & $1.9 \times 10^5$ & 40 & 620 & 2.1\\
Galaxy Cluster (c) & $9 \times 10^3$   & 600& 3300 & 174\\
\enddata
\label{tbl1}
\end{deluxetable}

\clearpage

\begin{deluxetable}{lccccccc}
\tablecolumns{7}
\tablewidth{0pc}
\tablecaption{Orbit parameters}
\tablehead{
\colhead{Orbit}&\colhead{$R_{a}$}&\colhead{$R_{p}$}&\colhead{$R_{0}$}&\colhead{$v(R_{a})$}
& \colhead{T}   & \colhead{$L_H$ }   & \colhead{$E_H$}\\
\colhead{}     & \colhead{(kpc)} & \colhead{(kpc)} & \colhead{(kpc)}   & 
\colhead{(km/s)}    &\colhead{(Gyr)}&\colhead{(kpc
km/s)}&\colhead{([km/s]$^2$)}
}
\startdata
Galaxy Cluster (c)   &  120   & 120    & 120      &  964       & 0.75 & 115680 &
$-4.65\times 10^5$\\
Compact Group (g)    &  100  & 35    & 69.2    &  265       & 0.60 & 26500 &
$-5.24\times 10^5$\\
\enddata
\tablecomments{ $R_0$ is the radius of the circular c-type orbit, or the
      radius of the circular orbit with the same energy as the
	    eccentric g-type orbit.}
\label{tbl2}
\end{deluxetable}

\clearpage

\begin{deluxetable}{lcccc} 
\tablecolumns{5} 
\tablewidth{0pc}
\tablecaption{Tidal radii of the galaxy models}
\tablehead{
\colhead{}&     \colhead{K or $\alpha$} & \colhead{Kee or
$\beta$} & \colhead{$\gamma$} & \colhead{Monopolar}\\} 
\startdata
c     & 22.69    &  15.28  & 13.75   &  15.28 \\ 
g-p  &  9.85    &   7.24  &  6.33  &   7.08 \\ 
g-R$_0$ & 15.78    &  10.56  &  9.85   &  10.80 \\ 
g-a  & 25.82    &  16.10  & 15.74  &  16.99 \\ 
\enddata
\label{tbl3}
\end{deluxetable}

\clearpage

\begin{deluxetable}{lcccccccccccccc}
\tablecolumns{15}
\tablewidth{0pc}
\tablecaption{Galaxy parameters}
\tablehead{
\colhead{}& \colhead{}& \colhead{}&\colhead{}&\multicolumn{3}{c}{g-p}&\colhead{}
&\multicolumn{3}{c}{g-$R_0$}&\colhead{}& \multicolumn{3}{c}{g-a}\\
\colhead{}& \colhead{c-Kee} & \colhead{c-K}&\colhead{}  & \colhead{$\alpha$} &
\colhead{$\beta$} & \colhead{$\gamma$}&\colhead{} & \colhead{$\alpha$} & 
\colhead{$\beta$} & \colhead{$\gamma$}&\colhead{} &
\colhead{$\alpha$} & \colhead{$\beta$} &\colhead{$\gamma$} }
\startdata
$W_o$                & 3.79& 3.88&& 3.53& 3.50& 3.41&&4.68& 4.60& 4.56&& 5.65& 5.59& 5.58\\
$\sigma_o$ (km/s)    & 273 & 263 && 309 & 317 & 327 &&248 & 256 & 259 && 209 & 215 & 216 \\
$t_{\rm dyn}$ (Myrs) & 120 & 220 && 63  & 40  & 32  &&128 & 69  & 63  && 267 & 132 & 127\\
\enddata
\label{tbl4}
\end{deluxetable}


\clearpage
\begin{thebibliography}{}

\bibitem[]{Athanassoula97} Athanassoula, E., Makino, J., \& Bosma, A. 1997, 
                         MNRAS, 286, 825
\bibitem[]{Barnes85} Barnes, J.E. 1985, MNRAS, 215, 517
\bibitem[]{Binney87} Binney, J., \& Tremaine, S. 1987, Galactic Dynamics
                   (Princeton: Princeton Univ. Press)
\bibitem[]{Bode93} Bode, P.W., Cohn, H.N., \& Lugger, P.M. 1993, ApJ, 416, 17
\bibitem[]{Bode94} Bode, P.W., Berrington, R.C., Cohn, H.N.,
                \& Lugger, P.M. 1994, ApJ, 433, 479
\bibitem[]{Boehringer97} Boehringer, H. 1997, in Dark Matter in Astro and 
                  Particle Physics (DARK '96), eds. H.V. Klapdor-Kleingrothaus,
                  \& Y. Ramachers, Y. (Heidelberg: World Scientific), 64
\bibitem[]{Brosche99} Brosche, P., Odenkirchen, M., \& Geffert, M. 1999, New
Astron., 4, 133
\bibitem[]{Chandra} Chandrasekhar, S. 1943, ApJ, 97, 255
\bibitem[]{DTGF98} Dom\'{\i}nguez-Tenreiro, R., \& G\'omez-Flechoso, M.A.
1998, MNRAS, 294, 465
\bibitem[]{Funato93} Funato, Y.,Makino, J., \& Ebisuzaki, T. 1993, PASJ, 45,
289
\bibitem[]{Garcia96} Garc\'{\i}a-G\'omez, C., Athanassoula, E., \& Garijo, A.
                   1996, A\&A, 313, 363
\bibitem[]{Gomez97} G\'omez-Flechoso, M.A. 1997, PhD thesis, Universidad
Aut\'onoma de Madrid
\bibitem[]{GFDT2000} G\'omez-Flechoso, M.A., \& Dom\'{\i}nguez-Tenreiro, R.
2000, in preparation (GD00)
\bibitem[]{gunn72} Gunn, J.E., \& Gott, J.R. 1972, ApJ, 176, 1
\bibitem[]{Heggie95} Heggie, D.C., \& Ramamani, N. 1995, MNRAS, 272, 317
\bibitem[]{Hernquist87} Hernquist, L. 1987, ApJSS, 64, 715
\bibitem[]{Innanen83} Innanen, K.A., Harris, W.E., \& Webbink, R.F. 1983, AJ,
88, 338
\bibitem[]{Jefferys76} Jefferys, W.H. 1976, AJ, 81, 983
\bibitem[]{Keenan81} Keenan, D.W. 1981a, A\&A, 95, 334; 1981b, A\&A, 95, 340 
\bibitem[]{Keenan75} Keenan, D.W., \& Innanen, K.A. 1975, AJ, 80, 290
\bibitem[]{King62} King, I.R. 1962, AJ, 67, 471
\bibitem[]{King65} King, I.R. 1965, AJ, 70, 376
\bibitem[]{King66} King, I.R. 1966, AJ, 71, 64
\bibitem[]{KGKK} Klypin, A.A., Gottlober, S., Kravtsov, A.V., \&
                 Khokhlov, A.M. 1999, ApJ, 516, 530
\bibitem[]{Kruskal62} Kruskal, D.   1962, J. of Math. Phys., 3, 806
\bibitem[]{Lee90} Lee, H.M. 1990, J. Korean Astron. Soc., 23, 97
\bibitem[]{Mateo98} Mateo, M. 1998, ARA\&A, 36, 435
\bibitem[]{Merritt85} Merritt, D. 1985, ApJ, 289, 18
\bibitem[]{Meylan97} Meylan, G., \& Heggie, D.C. 1997, A\&ARev, 8, 1
\bibitem[]{Meziane96} Meziane, K., \& Colin, J. 1996, A\& A, 306, 747
\bibitem[]{Michie63} Michie, R.W. 1963, MNRAS, 125, 127
\bibitem[]{Moore98} Moore, B., Lake, G. \& Katz, N. 1998, ApJ, 495, 139
\bibitem[]{Navarro96} Navarro, J.F., Frenk, C.S., \& White, S.D.M. 1996, ApJ,
                    462, 563
\bibitem[]{Oh92} Oh, K.S., Lin, D.N.C. 1992, ApJ, 386, 519
\bibitem[]{Ohal92} Oh, K.S., Lin, D.N.C., \& Aarseth, S.J. 1992, ApJ, 386, 506
\bibitem[]{Ponman96} Ponman, T.J., Bourner, P.D.J., Ebeling, H., \&  
         Boehringer, H. 1996, MNRAS, 283, 690
\bibitem[]{Spitzer87} Spitzer, L. 1987, Dynamical Evolution of Globular
                   Clusters (Princeton: Princeton Univ. Press) 
\bibitem[]{Weinberg94} Weinberg, M.D. 1994, AJ, 108, 1398
\end{thebibliography}
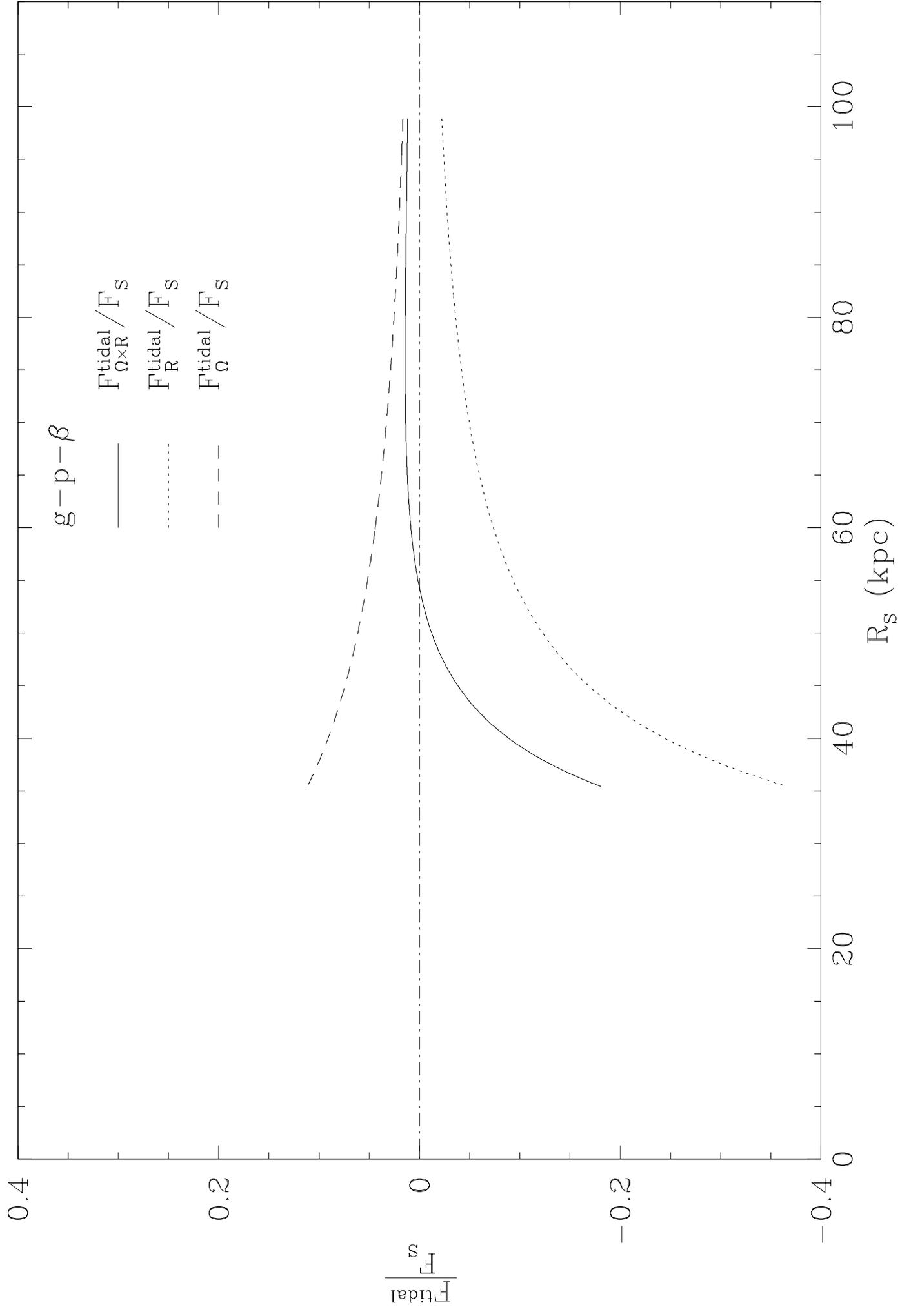
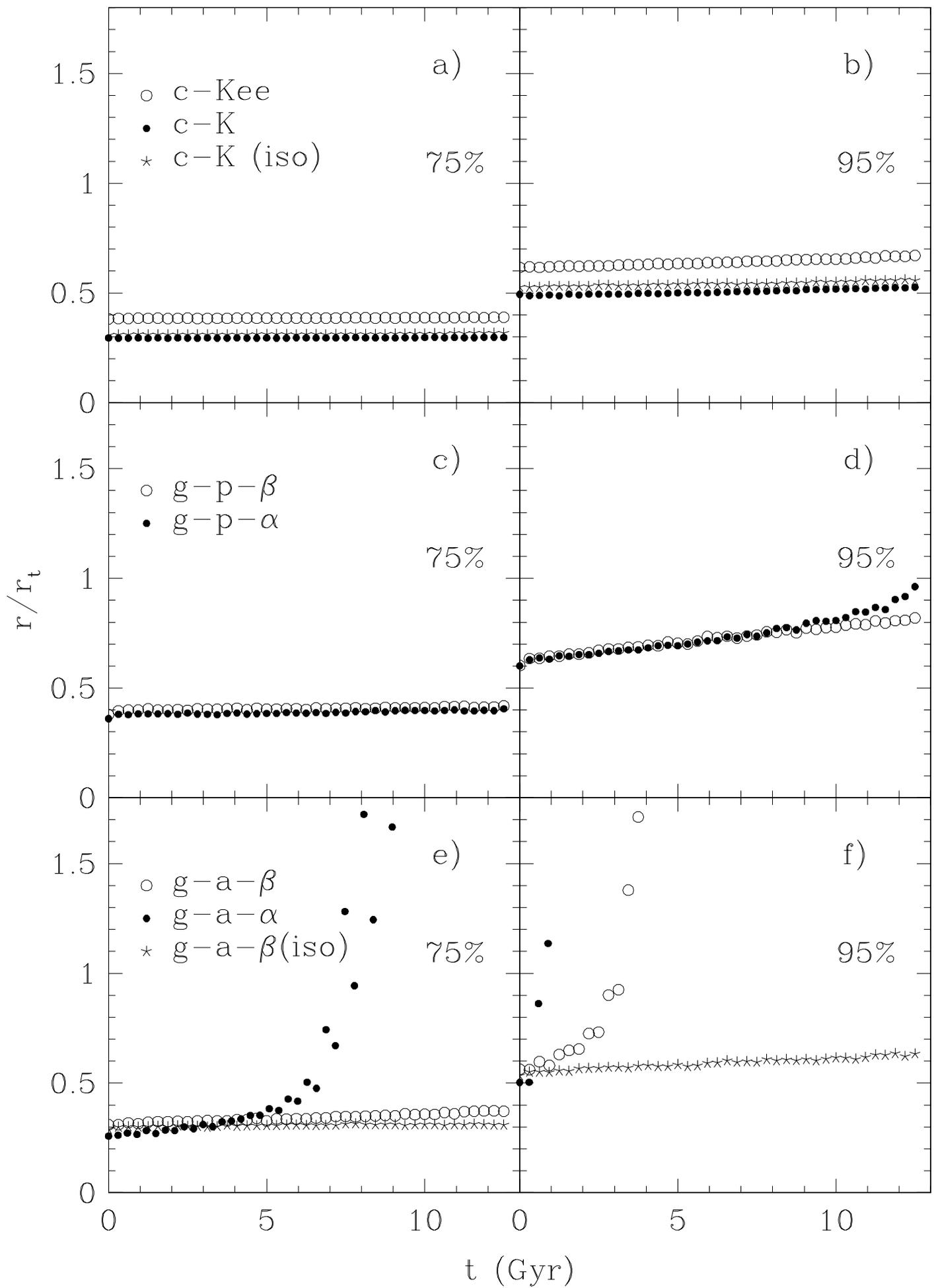
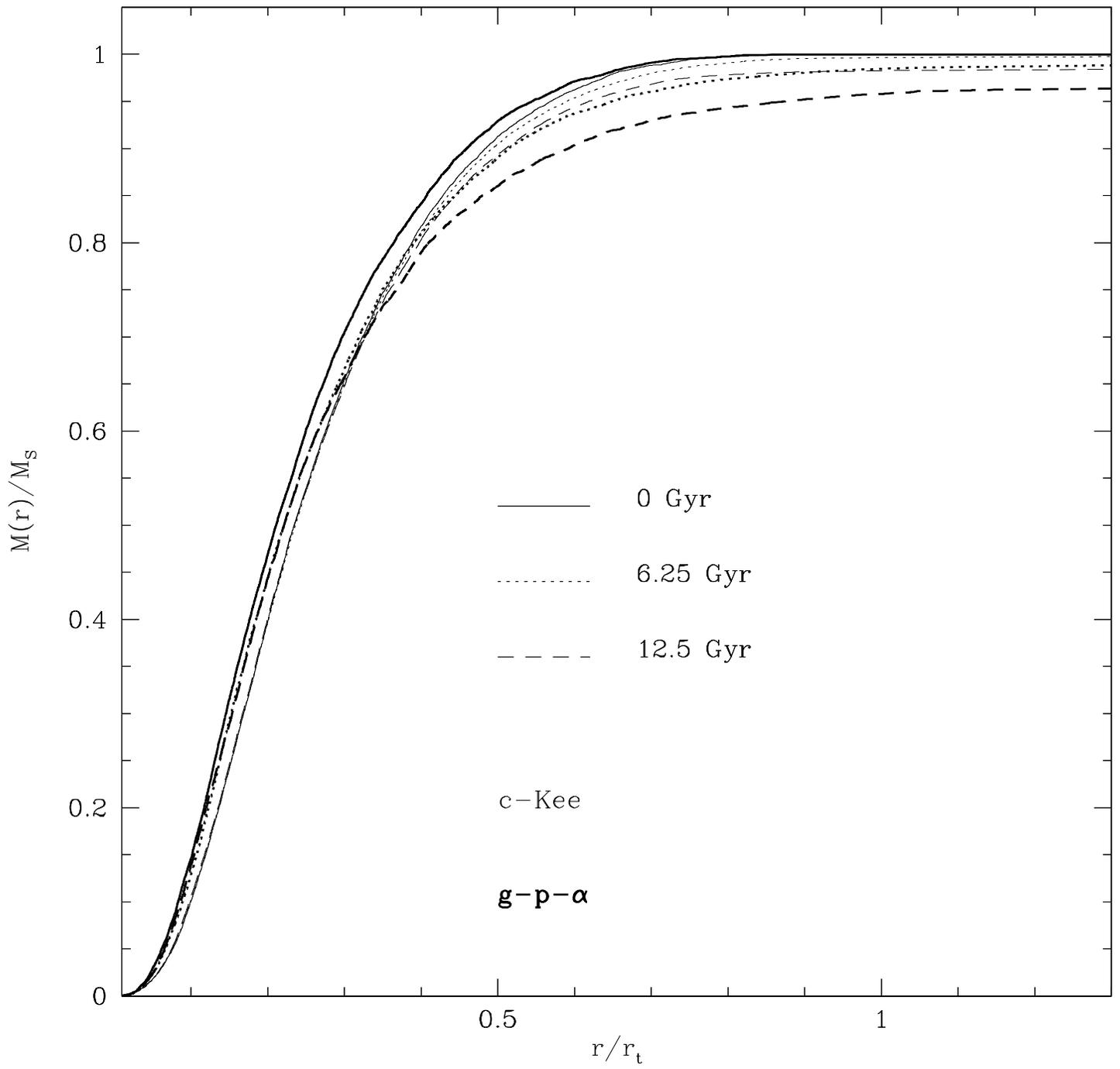
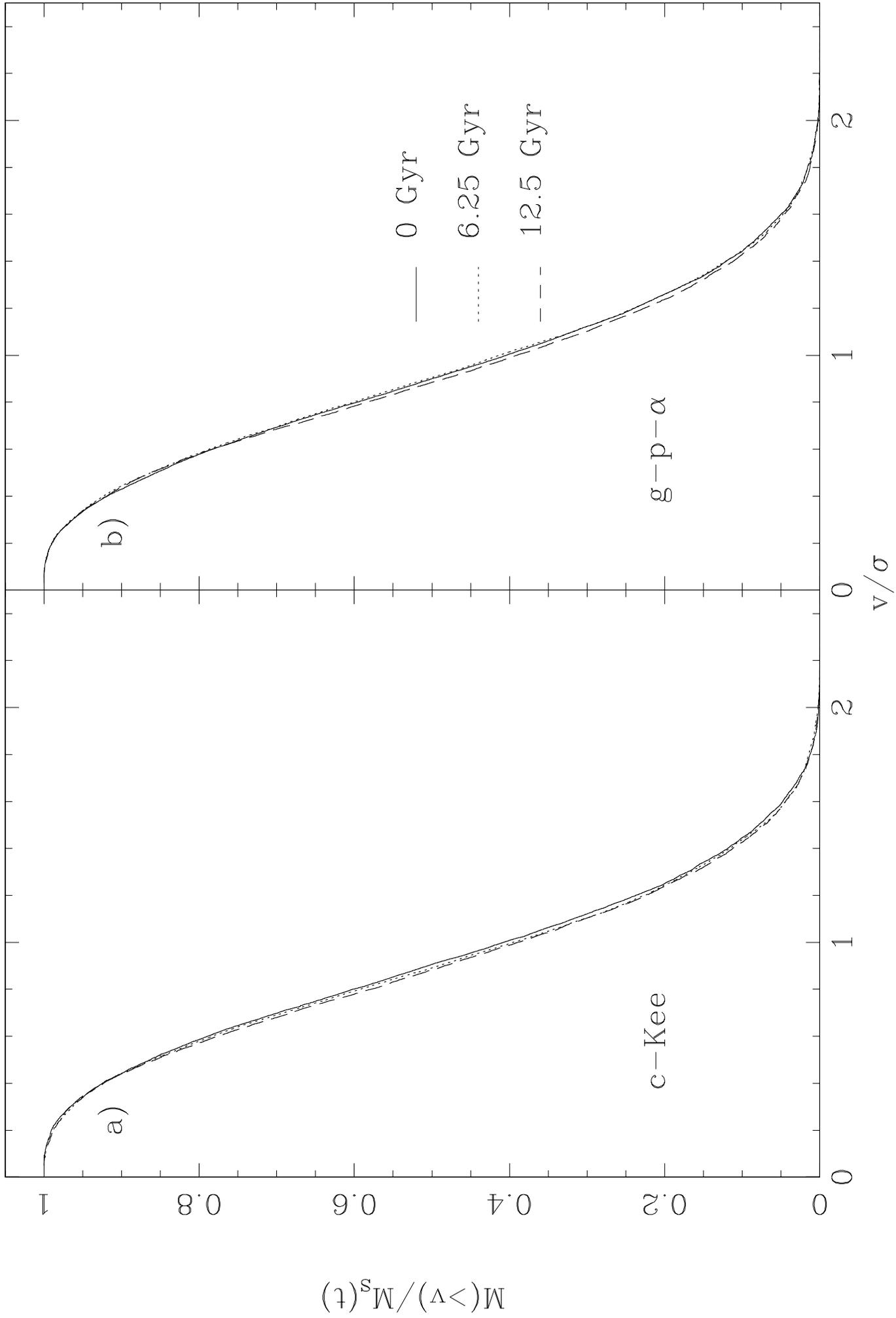
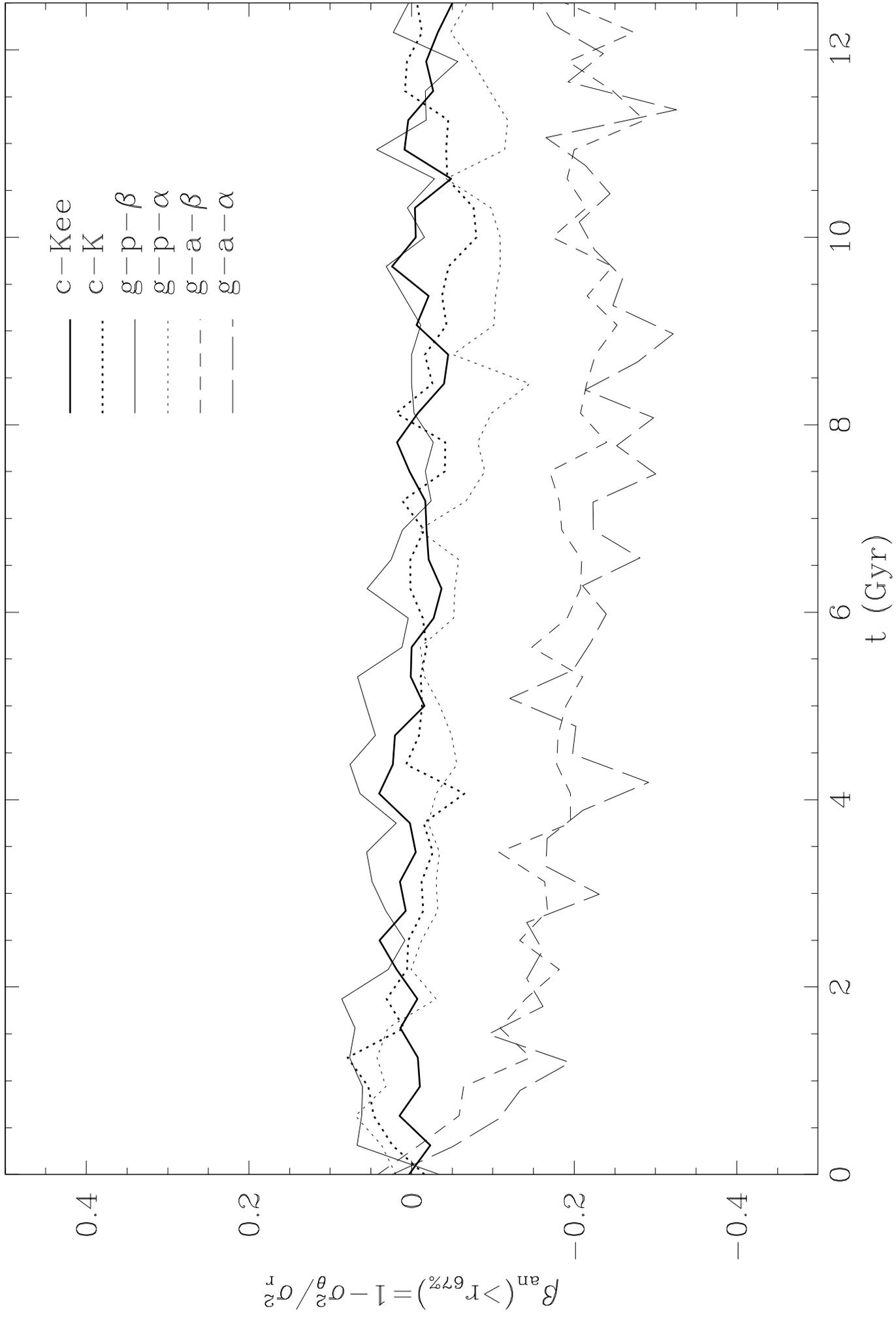
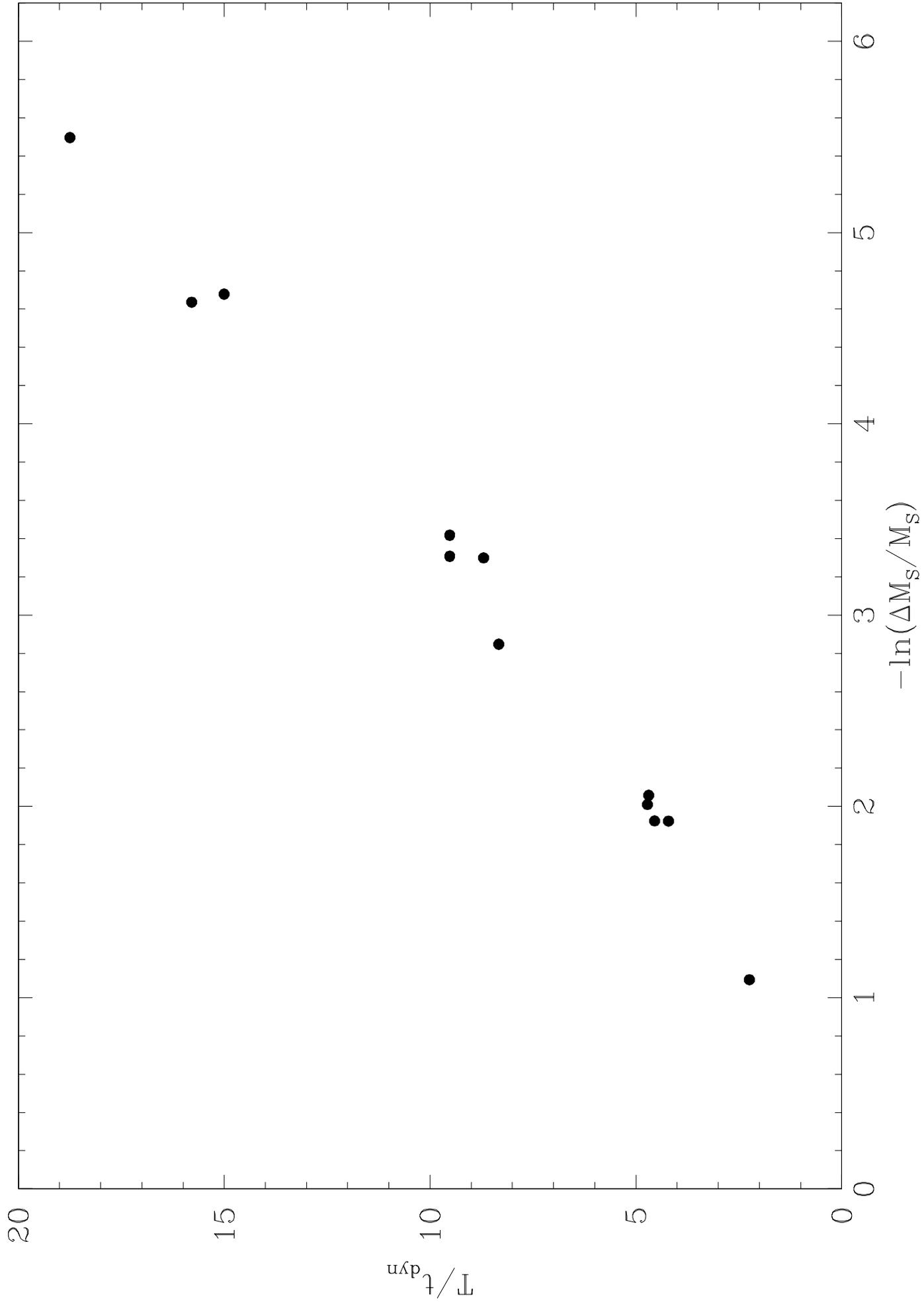
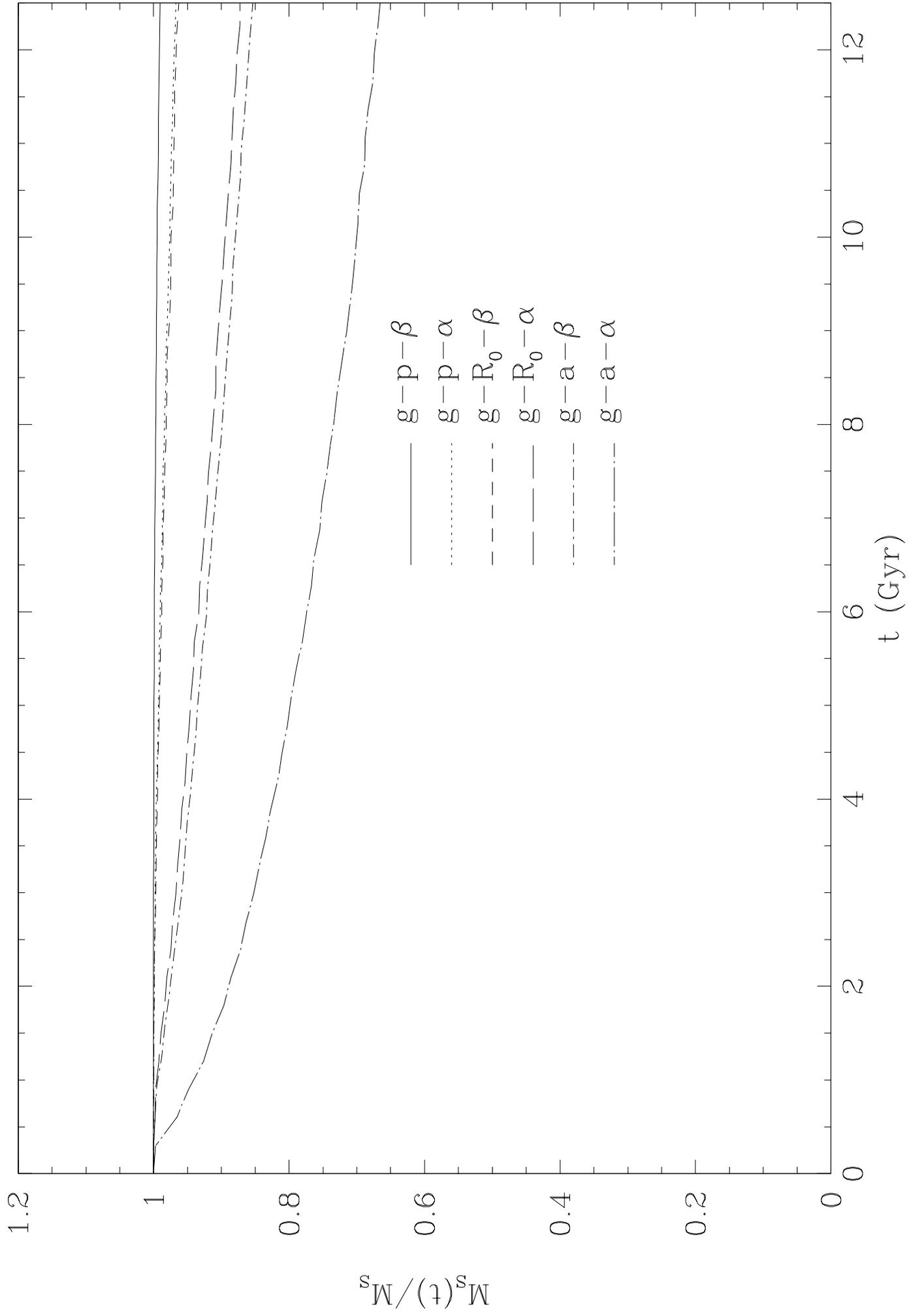
\end{document}